# Impact of external magnetic fields on STT-MRAM

Bernard Dieny[1], Sanjeev Aggarwal[2], Vinayak Bharat Naik[3], Sebastien Couet[4], Thomas Coughlin[5], Shunsuke Fukami[6,7], Kevin Garello[1], Jack Guedj[8], Jean Anne C. Incorvia[9], Laurent Lebrun[10], Kyung-Jin Lee[11], Daniele Leonelli[12], Yonghwan Noh[13], Siamak Salimy[10], Steven Soss[3], Luc Thomas[14], Weigang Wang[15], and, Daniel Worledge[16]

1. Univ. Grenoble Alpes, CEA, CNRS, SPINTEC, 38000 Grenoble, France
2. Everspin Technologies, 5670 W Chandler Blvd Suite 130, AZ 85226, USA
3. GlobalFoundries, 400 Stone Break Rd Extension, Malta, NY 12020, USA
4. Imec, 3001 Leuven, Belgium
5. Coughlin Associates, Atascadero, California, USA
6. Laboratory for Nanoelectronics and Spintronics, Research Institute of Electrical Communication, Tohoku University, 2-1-1 Katahira, Aoba-ku, Sendai 980-8577, Japan.
7. Center for Science and Innovation in Spintronics, Tohoku University, 2-1-1 Katahira, Aoba-ku, Sendai 980-8577, Japan.
8. NUMEM, Sunnyvale, California, USA
9. Dept. of Electrical and Computer Engineering, The University of Texas at Austin, Austin, Texas, USA
10. Hprobe, 4 Rue Irène Joliot Curie, 38320 Eybens, France
11. Department of Physics, Korea Advanced Institute of Science and Technology (KAIST), Daejeon, Korea.
12. Huawei Technologies R&D Belgium N.V., Leuven, Belgium
13. Netsol, Dongtan-daero 23-gil, Hwaseong-si, Gyeonggi-do, Republic of Korea.
14. Applied Materials, Santa Clara, CA 95054-3299 USA.
15. Department of Physics and Department of Electrical & Computer Engineering, the University of Arizona, Tucson, AZ 85721 USA.
16. IBM Almaden Research Center, San Jose, California, USA

## I. INTRODUCTION

This application note discusses the working principle of spin-transfer torque magnetoresistive random access memory (STT-MRAM) and the impact that magnetic fields can have on STT-MRAM operation. Sources of magnetic field and typical magnitudes of magnetic fields are given.

Based on the magnitude of commonly encountered external magnetic fields, we show below that magnetic immunity of STT-MRAM is sufficient for most uses once the chip is mounted on a printed circuit board (PCB) or inserted in its working environment. This statement is supported by the experience acquired during 60 years of use of magnetic hard disk drives (HDD) including 20 years of HDD with readers comprising magnetic tunnel junctions, 20+ years of use of magnetic field sensors as position encoders in automotive industry and 15+ years of use of MRAM. Mainly during chip handling does caution need to be exercised to avoid exposing the chip to excessively high magnetic fields.

MRAM stores data bits in magnetic tunnel junctions (MTJs). These are composed of two magnetic layers separated by an oxide tunneling barrier. One of the magnetic layers, called the free layer (or storage layer), has a switchable magnetization that can be switched between two stable directions, while the other layer, called the reference layer (or fixed layer), has a fixed magnetization direction. All STT-MRAM chips available today have their magnetization pointing perpendicular to the plane of the layers, up or down as indicated in Fig.1 [1]. This bistable behavior is realized by using magnetic layers designed to exhibit magnetic anisotropy normal to the plane of the layers, meaning that their magnetization has preferred out-of-plane orientation, either up or down.

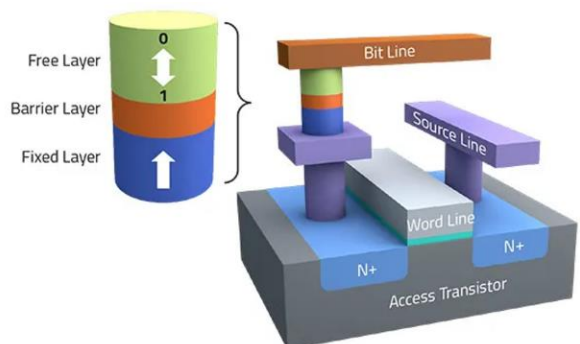

Fig.1. STT-MRAM cell containing a magnetic tunnel junction and a selection transistor. The magnetic tunnel junction is here represented just above the first metal level but is usually integrated in a higher metal layer.

The magnitude of this anisotropy is directly correlated with the stability of the magnetic configuration, which in turn determines the thermal stability of the stored bit and the current and energy required to switch between states.

When a current flows vertically through the tunnel barrier, the resistance of the tunnel barrier is low when the magnetizations of the storage and reference layers are parallel ($R_P$) and high when they are antiparallel ($R_{AP}$). The change of resistance between these two configurations is called the tunnel magnetoresistance (TMR) and is defined by: $\frac{\Delta R}{R_P} = \frac{R_{AP} - R_P}{R_P}$. With presently used MgO-based magnetic tunnel junctions, the TMR is of the order of 200 % meaning $\frac{R_{AP}}{R_P} \approx 3$. The binary information of the cell is stored in the magnetic configuration, either parallel or antiparallel.

An analogy can be drawn between a magnetic tunnel junction and a polarizer/analyzer pair in optics. In optics, the transmission of light through a polarizer/analyzer pair can be controlled by adjusting the angle between the polarizer and analyzer. Similarly, in magnetic tunnel junctions, the magnetic layers act as polarizer and analyzer for electron spins. The electron flow and therefore the device electrical resistance can be controlled by varying the angle between the magnetization of the storage and reference layers.

The MTJ is connected in series with a selection transistor, used to select that particular MTJ for reading or writing out of a rectangular array of MTJs.

During write operation, the storage layer magnetization is switched up or down by flowing a current of sufficient amplitude through the tunnel barrier, thanks to a phenomenon called spin transfer torque. To write in the parallel configuration, electrons flow from the reference layer to the storage layer. The tunneling electrons become spin-polarized by the reference layer, acquiring a spin orientation parallel to the magnetization of the reference layer. Upon entering the storage layer, their spins interact with the spins responsible for the storage layer's magnetization through a quantum interaction called exchange interaction. This interaction tends to align the storage layer's magnetization parallel to the injected spins, thereby aligning it parallel to the magnetization of the reference layer.

To write the antiparallel state, electrons flow from the storage layer to the reference layer. Electrons in the storage layer with spins antiparallel to the reference layer magnetization have a lower probability of tunneling compared to those with spins parallel to the reference layer magnetization. This results in an accumulation of spins antiparallel to the reference layer magnetization within the storage layer. The exchange interaction between these accumulated spins and the spins responsible for the storage layer's magnetization tends to induce an antiparallel alignment between the storage layer and reference layer magnetization.

In present STT-MRAM products, the write pulse is a few tens of nanoseconds long and the write voltage across the MTJ during write is $\approx 0.5$ V. At this voltage, write endurance can be quite high ($\approx 10^{12}$ cycles). It is possible to write faster but this comes at the expense of higher write voltages resulting in shorter endurance because the write voltage then gets closer to the dielectric breakdown voltage of the tunnel barrier.

During read operation, a significantly lower current than the write current (more than three times lower) is used to determine the resistance of the magnetic tunnel junction (low or high) without affecting the magnetization of the free layer. The chosen value of read current results from a trade-off between sufficiently fast read access time and low enough read disturb due to spin transfer torque associated with the read current. The value of the resistance indicates the junction magnetic configuration.

In standby, the stability of the free layer magnetization is determined by its coercivity, which is the external field required to switch the free layer magnetization. This coercivity is correlated with the magnetic layer anisotropy. The field strength (in tesla, T) is typically in the range of 0.2 T to 0.5 T in STT-MRAM chips depending on the MRAM target application (embedded FLASH (eFlash) or stand-alone). The fixed layer is designed to have a higher magnetic stability than the storage layer characterized by a coercivity of about 0.5 T to 0.6 T. The result is that STT-MRAM chips can withstand magnetic fields up to a few hundreds of mT in standby.

As for all magnetic devices, STT-MRAM exhibits some sensitivity to external magnetic fields, the most critical situation being during write at the lowest and highest specified operating temperature. In most cases, the lowest temperature is the most critical because higher write currents are required at low temperatures due to an increase in magnetic stability. If a magnetic field opposing the magnetization of the storage layer in the state to be written is present, the write current may be insufficient to successfully switch the magnetization of the storage layer.

In current products, STT-MRAM chips can operate without error in fields up to a few tens of mT to one hundred mT depending on MRAM target application (10 mT corresponding to 100 gauss in centimeter-gauss-seconds, or CGS units). Considering the rate at which magnetic field drops from field sources, as explained below, the magnetic immunity of STT-MRAM is sufficient for most uses once the chip is mounted on a PC board or inserted in its working environment. In extreme magnetic environments, either having a sufficient separation from the magnetic field source or improvement to magnetic immunity, such as magnetic field shielding, would be needed.

## II. Magnetic Fields, Magnitudes, Sources and Government Standards

Magnetic fields play essential roles in numerous everyday technologies, including cell phones, radio speakers, electrical transmission lines, disk drives, motors, televisions, household appliances, and medical devices like magnetic resonance imaging instruments. Despite their widespread presence, encountering magnetic fields exceeding a millitesla is uncommon.

The International Commission on Non-Ionizing Radiation Protection (ICNIRP) has established exposure limits at 0.42 mT (4.2 G) for occupational workers and 0.08 mT (0.8 G) for the general public. Studies conducted on household and work environments consistently demonstrate measurements below 0.1 mT (1 G) at distances greater than 16 cm (6.3 in) from various appliances and tools. For perspective, these levels are of similar magnitude as the Earth's magnetic field, around 0.05 mT (0.5 G). These values reflect typical human experiences and remain substantially lower than STT-MRAM's maximum field tolerance of a few tens of mT (a few hundred gauss) to one hundred mT depending on MRAM target application.

Additionally, the internationally accepted standard from the International Electrotechnical Commission (IEC), IEC-61000-4-8, regulates the magnitude of magnetic fields that devices can encounter during regular operation, with a maximum value of 1.26 mT (12.6 G), well below the STT-MRAM specification of a few tens or one hundred of mT. Despite the widespread utilization and generation of magnetic fields in everyday devices, exposure to significant field strengths are rare due to the rapid weakening of the field with the distance from its source. The primary sources of magnetic fields are current-carrying wires and permanent magnetic materials, both yielding fields of specific magnitudes determined by their geometries. Generating substantial magnetic fields at significant distances from the source proves challenging in practice.

### III. How Fields are Measured and Calculated

This section covers the scientific vocabulary related to magnetic fields. Refer to Table 1 for a breakdown of the different units used to quantify the strength of a magnetic field. There exist two systems of units for magnetic quantities: the modern International System of Units (SI) and the older CGS one. Magnetic induction, denoted by B, is used to describe the magnetic conditions within a magnetic material. When characterizing the magnetic environment in free space, as seen around an MRAM device, magnetic induction B is simply proportional to the magnetic field H. The relationship between B and H is more complicated within a magnetic material.

TABLE I
UNITS FOR MAGNETIC PROPERTIES

|  | SI Units | CGS Units |
|---|---|---|
| **Magnetic induction, B** | tesla (T) | gauss (G) |
| **Magnetic field, H** | amps/meter (A/m) | oersted (Oe) |

In free space, all four units are related as follows:

$$1 \text{ oersted} = 1 \text{ gauss} = \frac{10^3}{4\pi}\frac{A}{m} = 10^{-4} \text{ tesla} = 0.1 \text{ mT}$$

Magnetic fields are vector quantities, necessitating both magnitude and direction for full description. Various commercially accessible simulation software programs facilitate magnetic field calculations. Given input about the sources of the magnetic fields, these programs solve Maxwell's equations to calculate the magnetic field's magnitude and orientation. These sources may include permanent magnets or currents, as seen in devices like cell phone speakers, or current-carrying lines.

As discussed above for STT-MRAM field insensitivity, a few tens of mT (for example, 30 mT) of field would correspond to:

$$300 \text{ oersted} = 300 \text{ gauss} = 28.8 \frac{kA}{m} = 30 \text{ mT}$$

Magnetic fields can be easily measured using a Hall probe as illustrated in Fig. 2:

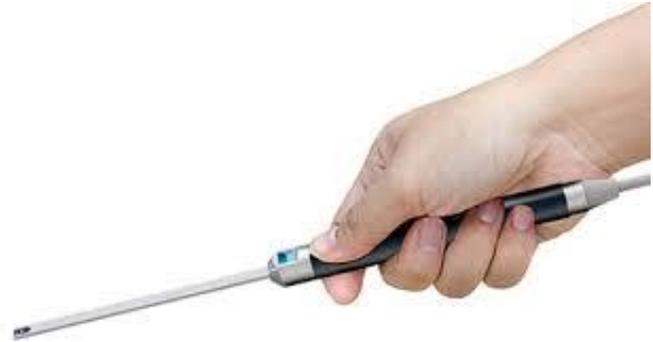

Fig.2. Illustration of use of a Hall probe sensor.

### IV. Examples of Calculated and Measured Magnetic Fields

We first consider a lengthy straight wire conducting 500 A, akin to the charging cable of an electric car with a cross-section measuring 1 centimeter. This scenario represents the upper limit of magnetic field generation for a current-carrying wire used for everyday applications. The formula utilized to compute the magnetic induction field in this instance is referred to as Ampere's Law:

$$B = \mu_0 H = \frac{\mu_0 I}{2\pi r} \quad \text{(in SI units, i.e. tesla) (a)}$$

In this expression, $\mu_0 = 4\pi \times 10^{-7} \text{T}/(A/m)$ is the vacuum permeability, $I$ is the current (in A), and $r$ is the radial distance from the wire center (in m). The direction of the magnetic field circulates around the wire as represented in Fig.3.

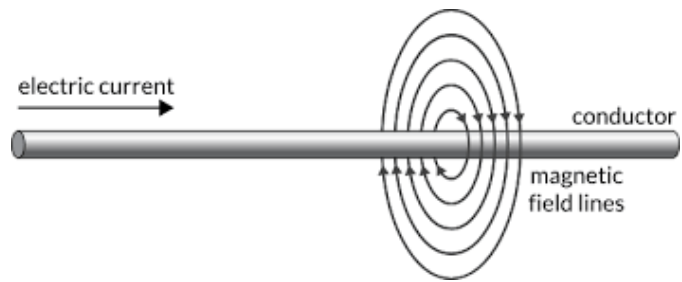

Fig.3. Magnetic field lines around an electrical wire.

As illustrated in Fig. 4, the strength of the magnetic induction field diminishes swiftly as radial distance from the center of the wire increases, never surpassing the STT-MRAM specification of a few tens of mT, even at the wire surface (i.e. at 0.5 cm of the wire center in this example).

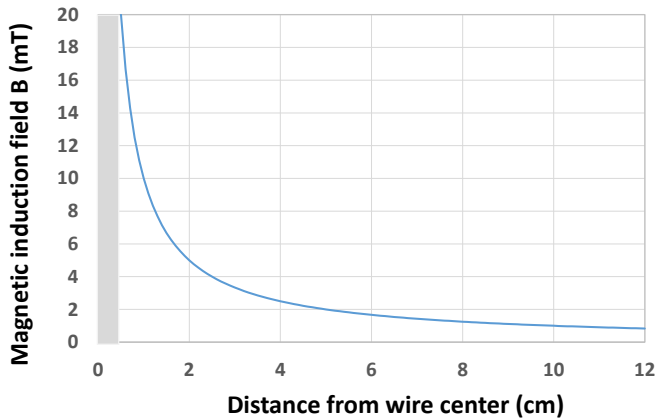

Fig.4: Magnetic induction field versus radial distance from the center of a wire carrying 500 A (Formula (a)). The length of the wire is here assumed to be much longer than 12 cm (infinite length limit) and frequency effects are not taken into account (assumption of dc current). The grey part represents an inaccessible zone inside the wire.

At a different scale, one can calculate the magnetic induction field at 1 µm distance from a wire carrying 1 mA. This field is 0.2 mT, again much smaller than the specified maximum field. In a different context, at ten meters from a 380 kV power line, the magnetic induction field is typically in the range of 1 µT to 10 µT, orders of magnitude too weak to have any impact on STT-MRAM.

We now consider measurements of the magnetic induction field created by a permanent magnet in the speaker of a cell phone. A magnetic strength of 9 mT was measured at the cellular phone speaker's surface. Within 5 mm of the surface, the field dropped below 1 mT, becoming negligible beyond 1 cm. All these values are well below the maximum field specification for STT-MRAM of a few tens of mT to one hundred of mT depending on the target application.

The magnetic field at the tip of a magnetic screwdriver is much larger; it can reach up to 100 mT. This magnitude exceeds the magnetic immunity threshold for most standalone chips today, and so care should be taken while handling standalone chips, depending on the safe magnetic field defined by the standalone chip manufacturer. However, the magnetic screwdriver would likely not cause errors in most embedded STT-MRAM with solder reflow retention during standby. Moreover, the field strength drops rapidly with distance, to below 25 mT at 1 mm distance.

Similarly, in direct contact with the strongest existing permanent magnets (made of NdFeB), the magnetic field can reach 700 mT; again, the field decreases rapidly with distance. As an example, Fig. 5 represents the magnetic field generated by a NdFeB magnet of dimensions 1 cm x 2 cm x 5 cm magnetized along the long axis (magnetization $\mu_0 M_s =$ 1.3 T). While the field exceeds 100 mT in very close proximity to the magnet (less than 1 cm away), it decreases rapidly with distance and gets below 30 mT at 2 cm from the magnet.

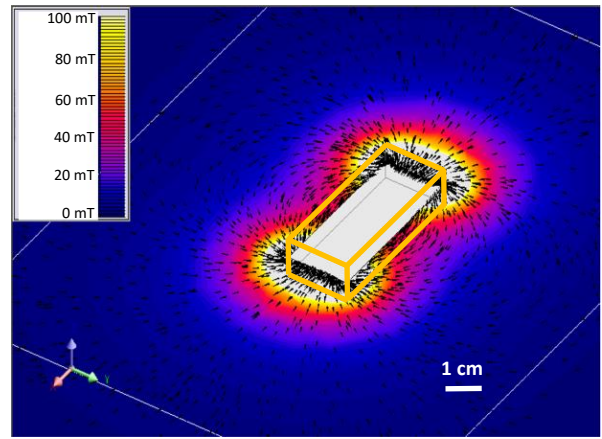

Fig. 5: Magnetic field generated by a strong permanent magnet of dimension 1 cm x 2 cm x 5 cm, magnetized along its length. The field intensity and field lines are represented in the XY median plane of the magnet, where the field is strongest.

A particular case of interest is that of cars, which may include several sources of magnetic fields. Magnetic fields have been measured at various places in a car as illustrated by Fig. 6. The measured magnetic fields are always below 1 mT except in direct contact with the surface of the phone holder magnet (if such a magnetic phone holder is used).

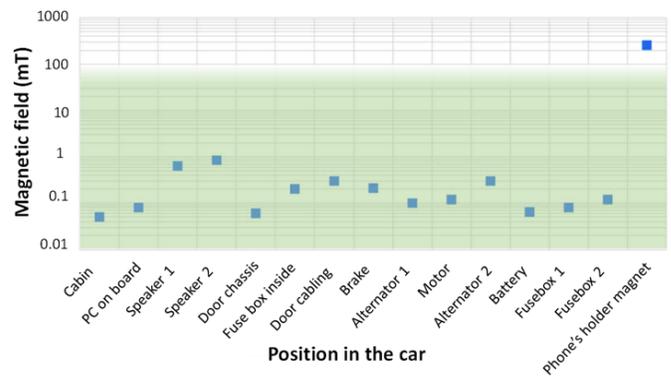

Fig.6: Magnetic field in a car measured at different positions (from [2]) The green background indicates the safe use zone.

In the vast majority of practical scenarios, the only instances one can come into sufficiently close proximity to the magnetic tunnel junctions to potentially disrupt the functioning of STT-MRAM is during handling and installation of the MRAM chip. Like handling electrostatically sensitive devices, special caution should be taken during MRAM chip manipulation considering the use of magnetic tools containing permanent magnets or materials that can become magnetized in the presence of a magnetic field and that can get within millimeters of the chip. Certain equipment and tools like pick and place machines and automated testing devices, commonly used during the PCB manufacturing may use rare earth magnets. It is recommended that MRAM devices be kept at least a few millimeters from such tools during PCB assembly operations if the field from the sources exceeds the safe magnetic field defined by the MRAM manufacturer. Non-magnetic versions of these tools are readily available and should be used if needed. Other examples of potentially magnetic tools include magnetic pencils/wands, tweezers, and wafer chucks. The handling environment can be easily checked for perturbing magnetic fields by using a Hall probe as illustrated in Fig. 2. In all cases, the magnetic field decreases rapidly with distance from the magnetic source.

Once the chip is placed on a PC board and inserted into its final environment in an industrial or automotive application, it is very unlikely that it will be exposed to a field exceeding the maximum specification of the MRAM chip due to the rapid decrease of field amplitude versus distance from practical magnetic field sources. In case of consumer and wearables applications, protective solutions can be used such as chip magnetic shielding, or guarding, by design, a distance to avoid any close presence of permanent magnets.

Numerous analytical, simulation, and measurement techniques are available to readily provide the assurances needed to adhere to the field immunity specification of the MRAM chip.

## V. STT-MRAM SENSITIVITY TO MAGNETIC FIELDS COMPARED WITH THAT OF OTHER MAGNETIC DEVICES

### A. Hard disk drives

Magnetic hard disk drives (HDD) have been used for data storage since 1956 (Fig.7).

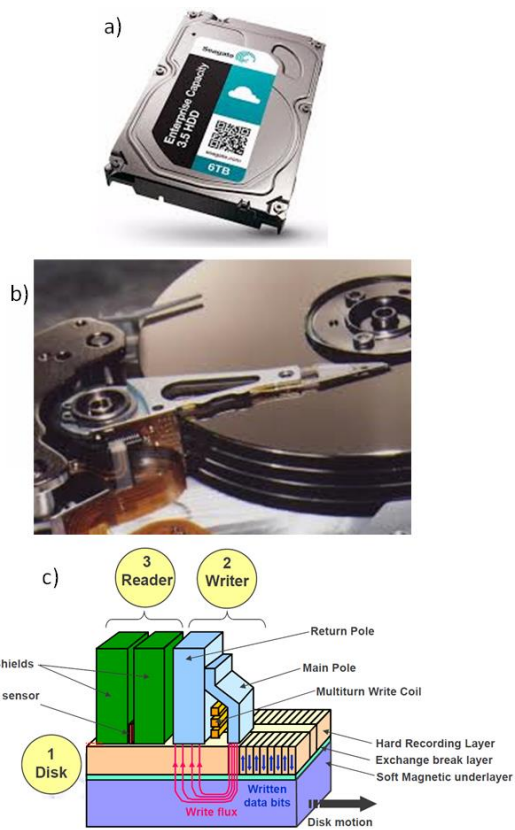

Fig. 7: Hard disk drives: a) HDD in its shielded enclosure, b) view of an open HDD with 4 stacked disks, c) working principle of HDD recording.

An HDD essentially consists of at least one disk whose surface is covered by a magnetic granular media and a write/read head flying above the spinning disk. The information is written by locally switching the magnetization of the grains up or down depending on the information to be written. Each grain's coercive field is of the order of 500 mT, about twice as large as that of the free layer in STT-MRAM. The grain magnetization is switched by the write head (writer part in Fig. 7c) which is a tiny electromagnet creating a magnetic field on the grains.

The information written in the media is read out by measuring the magnetic field just above the media at a distance $\approx$ 7 nm, corresponding to the flying height of the write/read head. This field is measured by a magnetoresistive sensor (reader part in Fig. 7c). It is of the order of a few mT only. This makes hard disk drives especially sensitive to external magnetic fields during read. As an example, the specification for maximum external field for the HGST CinemaStar Z7K500 (San Jose, CA) [3] at the enclosure surface is 1.5 mT in direct current mode, 0.5 mT at 60 Hz [4].

Despite these rather low values of maximum allowed field, HDDs with magnetic tunnel junction (MTJ) read sensors have been used ubiquitously with great success for about 20 years.

Given that the STT-MRAM specification permits a maximum external field about 20 times larger than for HDDs, there should be no concerns regarding magnetic immunity of STT-MRAM, once appropriate precautions have been taken during handling and installation as discussed above.

### B. Magnetic field sensors for automotive

Cars are equipped with numerous magnetic field sensors that function as linear or angular encoders. These sensors are utilized for various applications, such as detecting the position of the steering wheel, measuring wheel speed, monitoring chassis level, gauging fuel level, and determining throttle position, among others. Each encoder comprises two primary components: a fixed part, which is the magnetic field sensor itself, and a movable part, consisting of one or several permanent magnets attached to the moving mechanism. The relative position of these fixed and moving parts is determined by the magnetic field sensor, which measures the stray magnetic field generated by the moving permanent magnets. This stray field typically measures a few tens of millitesla (mT). If these sensors were exposed to magnetic fields exceeding a few tens of mT, they would malfunction. The long-term reliable use of these sensors in automotive applications suggests that STT-MRAM could also be deployed in cars without the risk of exposure to magnetic fields greater than a few tens of mT.

### C. Other MRAM technologies

The first type of MRAM commercially launched was Toggle MRAM (EVERSPIN, Chandler, AZ) in 2006, with a specified maximum allowed field for magnetic immunity between 2.5 mT and 15 mT, depending on product, in any direction [5].

These first MRAM chips are more sensitive to external magnetic field than current STT-MRAM chips because it is a field-written MRAM (not written by spin-transfer-torque) and the free layer of the magnetic tunnel junctions has much lower magnetic anisotropy. Toggle MRAM chips are magnetically shielded during packaging to improve their immunity to external magnetic fields [6].

To date, more than 150 million MRAM chips have been shipped with vanishingly small instances of magnetic-field-induced failures reported by consumers [7]. Best practices have been developed over time for handling and installing MRAM chips to ensure minimal contact with permanent magnets during chip soldering.

Several other companies are now producing STT-MRAM. In addition to standalone or discrete STT-MRAM products, major semiconductor foundries are producing embedded STT-MRAM within their back-end-of-line processing equipment. Everspin and Avalanche are producing standalone STT-MRAM, and Samsung, TSMC, and GlobalFoundries are producing embedded STT-MRAM.

In other MRAM technologies under research and development such as the ones based on spin–orbit torques and voltage controlled magnetic anisotropy, the coercivities of the free layer and the fixed layer of MTJs are of the same order of magnitude as that of STT-MRAM. Therefore they will function well within the same magnetic field range of a few tens of mT. In emerging MRAMs with antiferromagnetic free layers, the zero net magnetization of antiferromagnets will offer a superior magnetic immunity, which can further enhance the maximum operation field of MRAM by orders of magnitude.

Several commercial systems for testing chips electrically under exposure to magnetic fields are available to characterize the sensitivity of STT-MRAM chips to external magnetic field (provided by Hprobe (https://www.hprobe.com/) and Integral Solutions Int'l (https://www.us-isi.com/)). The results obtained with this test equipment are very consistent with theoretical expectations.

## VI. Conclusion

STT-MRAM chips have specifications regarding maximum external magnetic fields in the range of a few tens of mT to one hundred mT, depending on the MRAM offering for stand-alone or eFlash replacement. The current specifications far exceed typical values found in everyday environments by two orders of magnitude and surpass government standards by over 30x. These specifications are more than 40x higher than the specifications for widely-used hard disk drives with MTJ read sensors. In most applications, this magnitude should necessitate no design adjustments, and in special cases, require only minimal physical layout optimization.

Considering the rapid decrease of magnetic fields with distance from their sources, maintaining a separation of a few millimeters to one centimeter between MRAM devices and high magnetic field sources meets the specifications. Caution should be taken during chip handling, for example while mounting chips on a PC board. The magnetic environment can be easily checked by using a Hall probe setup. In addition, in very special cases where the risk of exposure to very high magnetic field exists, solutions can be used such as chip magnetic shielding and guarding, by design, a distance to avoid any close presence of permanent magnets.

- Daniel R. Symalla, 2022, Everspin, application note on magnetic immunity of EMxxLX devices, 07/30/2024: https://www.everspin.com/file/158196/download